\documentclass{emulateapj}
\newcommand{\mum}{\ifmmode{\rm \mu m}\else{$\mu$m}\fi}
\newcommand{\iras}{{\it IRAS}}
\newcommand{\iso}{{\it ISO}}

\begin{document}

\title{The unusual hydrocarbon emission from the early carbon star 
HD 100764:  The connection between aromatics and aliphatics}

\author{
G.~C.~Sloan\altaffilmark{1},
M.~Jura\altaffilmark{2},
W.~W.~Duley\altaffilmark{3},
K.~E.~Kraemer\altaffilmark{4},
J.~Bernard-Salas\altaffilmark{1},
W.~J.~Forrest\altaffilmark{5},
B.~Sargent\altaffilmark{5},
A.~Li\altaffilmark{6},
D.~J.~Barry\altaffilmark{1},
C.~J.~Bohac\altaffilmark{5},
D.~M.~Watson\altaffilmark{5},
\& J.~R.~Houck\altaffilmark{1}
%A. Scott\altaffilmark{3,7},
}
\altaffiltext{1}{Cornell University, Astronomy Department, 
  Ithaca, NY 14853-6801; sloan@isc.astro.cornell.edu, 
  jbs@isc.astro.cornell.edu, don@isc.astro.cornell.edu, jrh13@cornell.edu}
\altaffiltext{2}{Department of Physics and Astronomy and Center 
  for Astrobiology, University of California, Los Angeles, CA 
  90095-1562; jura@astro.ucla.edu}
\altaffiltext{3}{Department of Physics and Astronomy, University of Waterloo, 
  200 University Avenue West, Waterloo, ON N2L 3G1, Canada; 
  wwduley@uwaterloo.ca}
\altaffiltext{4}{Air Force Research Laboratory, Space Vehicles
  Directorate, 29 Randolph Rd., Hanscom AFB, MA 01731;
  kathleen.kraemer@hanscom.af.mil}
\altaffiltext{5}{Department of Physics and Astronomy, University 
  of Rochester, Rochester, NY 14627-0171; forrest@pas.rochester.edu,
  bsargent@pas.rochester.edu, cbohac@pas.rochester.edu, dmw@pas.rochester.edu}
\altaffiltext{6}{Department of Physics and Astronomy, University of 
  Missouri Columbia, MO 65211, lia@missouri.edu}
%\altaffiltext{7}{COM DEV International, 1725 Woodward Drive, 
%  Ottawa, ON K2C 0P9, Canada}

\slugcomment{Accepted in ApJ}

%\slugcomment{PREPRINT}

\begin{abstract}

We have used the Infrared Spectrograph (IRS) on the {\it 
Spitzer Space Telescope} to obtain spectra of HD 100764, an 
apparently single carbon star with a circumstellar disk.  The 
spectrum shows emission features from polycyclic aromatic
hydrocarbons (PAHs) that are shifted to longer wavelengths 
than normally seen, as characteristic of ``class C'' systems 
in the classification scheme of Peeters et al.  All seven of 
the known class C PAH sources are illuminated by radiation 
fields that are cooler than those which typically excite PAH 
emission features.  The observed wavelength shifts are 
consistent with hydrocarbon mixtures containing both aromatic 
and aliphatic bonds.  We propose that the class C PAH spectra 
are distinctive because the carbonaceous material has not 
been subjected to a strong ultraviolet radiation field,
allowing relatively fragile aliphatic materials to survive.

\end{abstract}

\keywords{ circumstellar matter --- stars:  carbon }

\section{Introduction} % Sec. 1.0

Complex hydrocarbons are found in a wide variety of 
astrophysical environments.  \cite{gfm73} first detected
what became known as the unidentified infrared (UIR) 
emission features, the strongest of which are at 3.3, 6.2, 
7.7--7.9, 8.6, 11.3, and 12.7~\mum.  \cite{lp84} and 
\cite{atb85} identified the carrier of these features as
polycyclic aromatic hydrocarbons (PAHs).  The debate over 
PAHs as the carrier of the UIR features raged for many 
years, and while the identification now seems firm 
\cite[e.g.][]{ahs99}, many questions still remain over 
the origin, evolution, and exact composition of astrophysical 
hydrocarbons.

\cite{pee02} used data from the Short Wavelength Spectrometer 
(SWS) on the {\it Infrared Space Observatory} (\iso) to group 
PAH spectra into three classes based on the positions of peak 
emission and profiles of the C--C skeletal deformation modes 
within the PAHs at 6.2 and 7.7--7.9~\mum.  The ``class A'' 
sources display features at 6.22 and 7.65~\mum\ and generally 
include H II regions, reflection nebulae, galaxies, and 
Herbig AeBe (HAeBe) stars still embedded in the regions where
they formed.  The ``class B'' sources, which typically 
include planetary nebulae and isolated HAeBe stars, show 
these same features, but shifted to 6.26 and 7.85~\mum.  
\cite{pee02} also identified two ``class C'' sources 
including the well-known Egg Nebula (AFGL 2688) which are in 
transition from the asymptotic giant branch (AGB) to the 
planetary nebula phase.  These two sources show emission at 
6.26~\mum, no emission near 7.65~\mum, and a broad feature 
centered at $\sim$8.2~\mum.  It remains unclear how these 
different classes of PAHs arise in different astrophysical 
environments.

Observations with the Infrared Spectrograph 
\citep[IRS;][]{hou04} on the {\it Spitzer Space Telescope} 
\citep{wer04} have added to the small sample of class C PAH 
spectra.  \cite{kra06}  reported the detection of a class C 
PAH spectrum from MSX SMC 029, a transition object in the 
Small Magellanic Cloud (SMC).  \cite{jur06} detected a class 
C PAH spectrum from HD 233517, an apparently single red giant 
(K2 III) with a substantial infrared excess which most likely 
arises from a circumstellar disk \citep{jur03}.  This system 
is remarkable because some of the hydrocarbons were probably 
synthesized from the oxygen-rich disk.  

% \section{HD 100764}

HD 100764 is one of a handful of red giants (luminosity class 
III) with large infrared excesses characteristic of orbiting 
disks.  While most of these stars are oxygen-rich like 
HD 233517, HD 100764 is carbon-rich.  As part of a 
program to better understand these stars, we obtained IRS 
data for HD 100764.  We have found that its IRS spectrum 
reveals low-contrast class C PAH features not detectable in 
the previous ground-based spectroscopy.  In this paper, we 
describe the PAH spectrum of HD 100764 in detail and relate 
it to other class C PAH sources.
% and present a model to explain the properties of these systems.

HD 100764, also known as CCS 1886 and CGCS 3066, is 
classified as an early carbon star, R2 \citep{san44} or 
C1,1 \citep{yam72}.  Using optical spectroscopy, \cite{dom84} 
found that the central star has a temperature of 4850 K, is 
iron poor, and has an enhanced $^{13}$C/$^{12}$C ratio, 
making this carbon star one of the uncommon J stars.  The 
Hipparcos parallax of 2.78 ${\pm}$ 1.18 mas yields an 
uncertain distance of $\sim$360 pc and thus a total 
luminosity of 73 L$_{\odot}$ \citep{par91}, indicating that 
the star is a first-ascent red giant.

\cite{le90} identified HD 100764 as a potential 
silicate/carbon star because of its red \iras\ [12]$-$[25] 
color as measured by the {\it Infrared Astronomy Satellite} 
(\iras).  As a consequence, it was observed twice using 
ground-based mid-infrared spectrometers over a decade ago.  
\cite{lev92} obtained an 8--14~\mum\ spectrum at the
Wyoming Infrared Observatory (WIRO) with a resolution 
($\lambda/\Delta\lambda$) of $\sim$50 and a signal/noise 
(S/N) ratio of $\sim$5.  While they could rule out the 
presence of silicate dust emission at 10~\mum\ in the 
spectrum, they could say little more.  \cite{ski94} obtained 
a higher quality spectrum with CGS3 at the United Kingdom 
Infrared Telescope (UKIRT), also saw no spectral features 
from dust, and concluded that the infrared emission was 
dominated by amorphous carbon dust in a circumstellar disk.
The disk explains why the strong infrared excess does not 
appreciably redden the optical spectrum.  

Although uncertain, it is plausible that this disk formed 
when HD 100764 became a red giant and engulfed a companion.
The luminosity of HD 100764 is consistent with other R
stars with measured parallaxes, which \cite{alk98} have
determined are first-ascent red giants, not AGB stars. 
\cite{mcc97} observed that R stars do not have companions
and suggested that their carbon-rich nature and the lack of 
a companion could both be explained by coalescence of an
expanding giant and its companion.

\section{The infrared spectra} % Sec. 2.0

\subsection{Observations and analysis} % Sec. 2.1

We observed HD 100764 with both the Short-Low (SL) and 
Long-Low (LL) modules of the IRS in standard staring mode on 
2006 January 14.  Because the source is bright for the IRS, 
we used the shortest possible ramps in both modules, 6-second 
ramps which provide only four samples.  This source was 
observed independently with the Short-High (SH) and Long-High 
(LH) modules in {\it Spitzer} program 278 (AOR key 19481344), 
and we include those data here with permission of the P.I., 
M. Werner.  % 3235

We followed the standard calibration method used at Cornell.  
We started with the S14 pipeline output from the {\it 
Spitzer} Science Center (SSC) and subtracted a sky image from
each image before extracting a spectrum.  The high-resolution 
observations include separate sky images obtained close to 
the target.  For SL, the sky image for one spectral aperture 
is obtained when the source is in the other aperture (i.e. 
aperture differerences:  SL2$-$SL1 and SL1$-$SL2, where 
``SL2'' means ``SL Order 2'').  For LL, we simply differenced 
the images with the source in the same aperture, but in the 
other nod position.  For a source as bright as HD 100764, the 
primary benefit of background removal is the subtraction of 
most of the rogue pixels.  These pixels have temporarily 
increased dark currents and are the primary source of noise 
in images with bright sources.  We applied {\rm 
imclean.pro}\footnote{Available from the SSC as {\rm 
irsclean.pro}.} to replace remaining rogue and bad pixels.  

Spectra were extracted from the individual images using 
the SSC-defined apertures and the routines {\rm profile}, 
{\rm ridge}, and {\rm extract}, which are available with the 
{\rm SPICE} software package ({\it Spitzer} IRS Custom
Extraction).   To calibrate to flux-density
units, we used a spectral correction generated from IRS 
observations of standard stars, HR 6348 (K0 III) for SL, and 
HR 6348, HD 166780 (K4 III) and HD 173511 (K5 III) for LL 
\citep[see ][ for more details]{slo07}.  The high-resolution 
data were calibrated using $\xi$ Dra (K2 III).  Error bars 
are based on the standard deviation of the spectrum in the 
two separate nods in each aperture.  The resulting spectrum 
has a S/N ratio which is typically between 200 and 400.

We applied scalar multiplicative corrections to the segments 
to remove discontinuities and align them upwards to the 
presumably best-centered spectral segment, in this case LL2.  
This step required corrections of 9\% to SL2, 2\% to SL1, and 
1\% to LL1.  These corrections indicate the accuracy of the 
pointing of the telescope and are consistent with typical
observations.  Finally, we combined the bonus order with 
valid data in the overlapping orders and removed all data 
outside the ranges of trustworthy signal.  Table 1 gives the 
wavelength ranges used for each low-resolution order.

% Table 1 goes here

\begin{deluxetable}{ccc} % Table 1
\tablecolumns{3}
\tablewidth{0pt}
\tablenum{1}
\tablecaption{Low-resolution wavelength ranges}
\tablehead{
  \colhead{order} & \colhead{SL (\mum)} & \colhead{LL (\mum)} }
\startdata
 second & 5.10--7.59  & 14.20--21.23 \\
 bonus  & 7.73--8.39  & 19.28--21.23 \\
 first  & 7.59--14.33 & 20.46--37.00 \\
\enddata
\end{deluxetable}

\begin{figure} % Fig. 1
\includegraphics[width=3.5in]{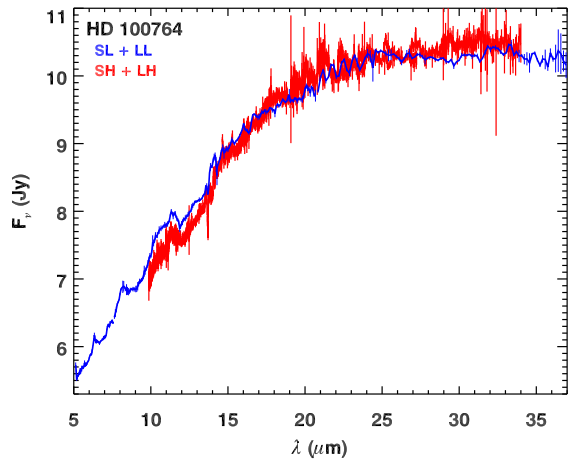}
\caption{The spectrum of HD 100764 with the low-resolution
modules (blue or black) and with the high-resolution 
modules (red or gray).}
\end{figure}

The flux density from HD 100764 in the 5--37~\mum\ spectral 
region ranges between 5 and 11 Jy, and we have therefore 
assessed the  saturation and non-linearities in the response 
function, especially in LL.  The S14 pipeline has a greatly 
improved correction for these problems, and it produces much 
better spectra compared to previous pipeline versions.  
Figure 1 presents the full spectrum of HD 100764, both our 
low-resolution data and also the high-resolution data 
obtained by M. Werner et al.  A comparison of the two spectra 
reveals no evidence for any residual artifacts from the
saturation and linearity corrections in our low-resolution 
spectrum.  One artifact that does remain is the fringing 
from the LL1 filter apparent between 21 and 25~\mum.  We
do not attempt a correction because the analysis in this
paper concentrates on the features at shorter wavelengths.
For this paper, we focus on the
low-resolution spectrum from 5 to 14~\mum.

Superimposed on the continuum produced by the circumstellar 
disk are emission features at wavelengths near those 
associated with PAHs and a broad emission feature centered 
at 10.6~\mum.  To isolate the PAH-related features we 
subtracted a spline-fitted continuum, as Figure 2 shows.  
This spline passes through and includes the broad feature 
peaking at 10.63~\mum.  We examine this feature separately 
in \S 2.5.  The spectrum of HD 100764 also shows a narrow 
absorption band from acetylene (C$_2$H$_2$) at 13.7~\mum.  
When this absorption band appears with PAH emission features, 
it is only in the rare class C PAH spectra \citep{kra06}.

The low-contrast features in the 8--14~\mum\ range escaped 
previous detection.  The 8~\mum\ feature has a contrast of 
5\% but lies at the edge of the ground-based window, and the 
feature near 11.3~\mum\ has a contrast of only 3\%, too small 
to be seen even in the spectrum by \cite{ski94}, which has a 
S/N ratio of $\sim$15 at those wavelengths.  

Figure 2 ({\it top}) plots the spectra from the two nod 
positions separately.  Systematic errors in the spectra have 
their greatest impact on low-contrast features, and they 
often show up as differences between the nods.  Some detailed 
differences in the PAH features are apparent, most notably at 
$\sim$ 8.5~\mum, but these differences are not large enough 
to have a significant quantitative effect on our conclusions.  
Figure 2 ({\it bottom}) presents the continuum-subtracted 
spectrum from HD 100764, along with line segments used to 
extract the strength and position of each PAH emission 
feature, as described in \S 2.3.

% KK - can we quantify here, e.g. lamba_c?

\begin{figure} % Fig. 2
\includegraphics[width=3.5in]{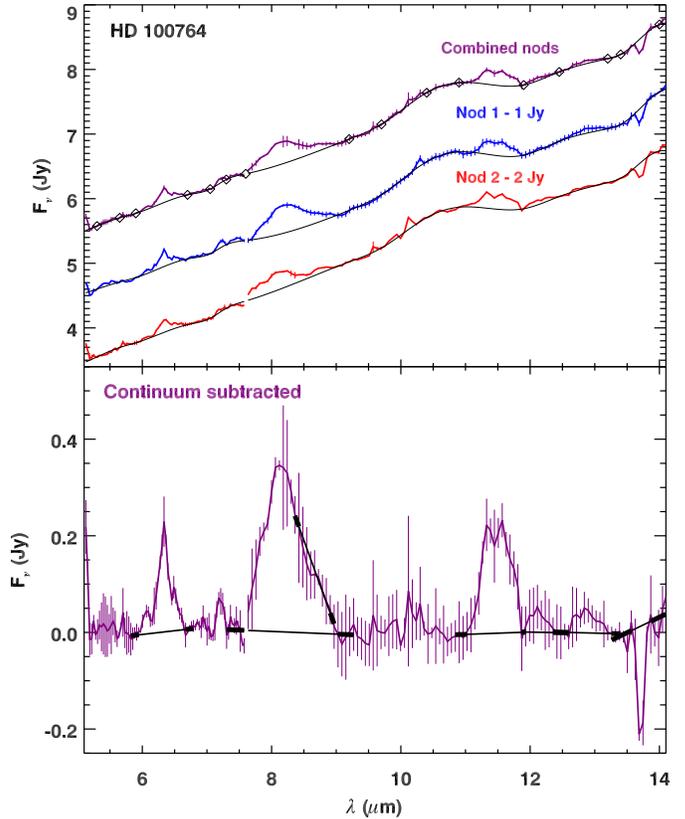}
\caption{The spectrum of HD 100764.  The top panel shows
the coadded spectrum and the spectra from the individual nod 
positions (offset by 1 and 2 Jy, respectively).  The thin 
solid curve is a spline fit to the data at wavelengths marked 
with diamonds.  These points lie outside of the PAH features 
and C$_2$H$_2$ absorption band at 13.7~\mum.  The bottom panel 
shows the resulting continuum-subtracted spectrum, along with 
the error bars and the line segments fit to extract the 
strengths and centers of the spectroscopic features.  The 
line segments are thick in the wavelength regions used to fit 
the continuum and thin where the feature is integrated.  The 
equivalent width of the 13.7~\mum\ acetylene band is measured 
before any continuum subtraction.}
\end{figure}

\subsection{The comparison sample} % Sec. 2.2

\begin{deluxetable*}{llrllll} % Table 2
\tablecolumns{7}
\tablewidth{0pt}
\tablenum{2}
\tablecaption{HD 100764 and the comparison sample}
\tablehead{
  \colhead{ } & \colhead{ } & \colhead{ } & \colhead{ } & 
  \colhead{Spectral} & \colhead{$T_{eff}$} & \colhead{PAH} \\
  \colhead{Target} & \colhead{Instrument} & \colhead{AOR key or TDT} & 
  \colhead{Object type\tablenotemark{a}} & \colhead{type\tablenotemark{a}} & 
  \colhead{(K)\tablenotemark{a}} & \colhead{class} }
\startdata
HD 100764       & IRS & 16262656          & red giant       & 
  C1,1    &  4850   & C \\
IRAS 13416-6243 & SWS & 62803904          & post-AGB        & 
  G1 I    &  5440   & C \\
AFGL 2688       & SWS & 33800604 35102563 & post-AGB        & 
  F5 Ia   &  6520   & C \\
MSX SMC 029     & IRS & 10656256          & post-AGB        & 
  \nodata & \nodata & C \\
HD 233517       & IRS &  3586048          & red giant       & 
  K2 III  &  4475   & C \\
SU Aur          & IRS &  3533824          & T Tauri         & 
  G1      &  5945   & C \\
HD 135344       & IRS &  3580672          & Herbig AeBe     & 
  F4 Ve   &  6590   & B/C \\
HD 44179        & SWS & 70201801          & post-AGB        & 
  A0 III  &  9520   & B \\
NGC 1333 SVS 3  & SWS & 65902719          & reflection neb. & 
  B6      & 14000   & A \\
\enddata
\tablenotetext{a}{See \S 3.1 for details and references.}
\end{deluxetable*}

\begin{figure} % Fig. 3
\includegraphics[width=3.5in] {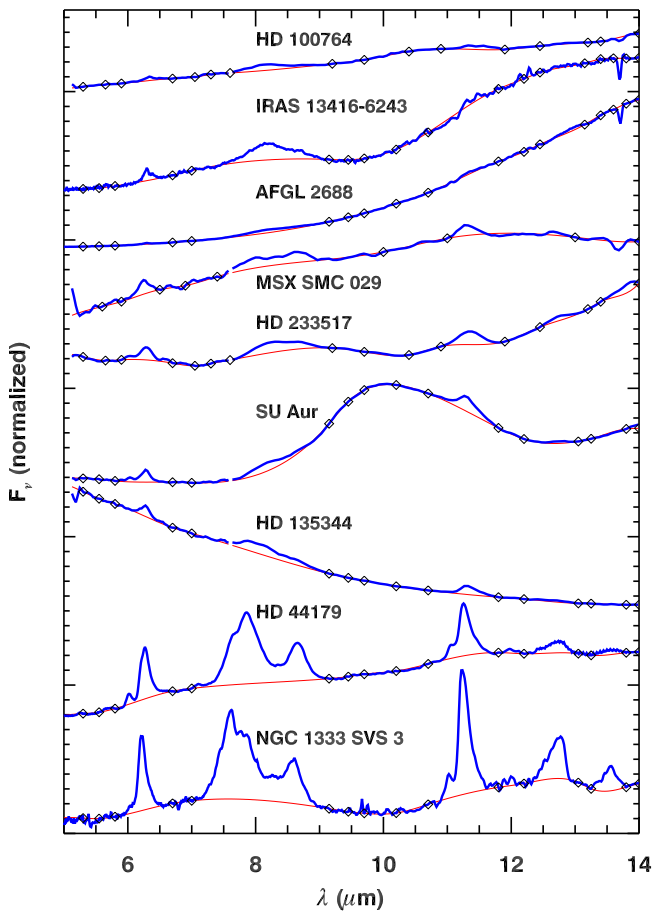}
\caption{Infrared spectra of several sources showing
class C PAH features, along with spectra of HD 44179
(class B) and NGC 1333 SVS 3 (class A).  The spectra of 
IRAS 13416, AFGL 2688, NGC 1333 SVS 3, and HD 44179 are 
from the \iso/SWS database \citep[as processed 
by][]{slo03}; the remaining spectra are from the IRS.  
The thin lines show the continua fit to the spectra, based 
on a spline through the wavelengths marked with diamonds.
The resulting continuum-subtracted PAH spectra appear 
in Fig. 4.  
%The letters in parentheses give the class
%of PAH emission, as defined by \cite{pee02}
}
\end{figure}

\begin{figure} % Fig. 4
\includegraphics[width=3.5in]{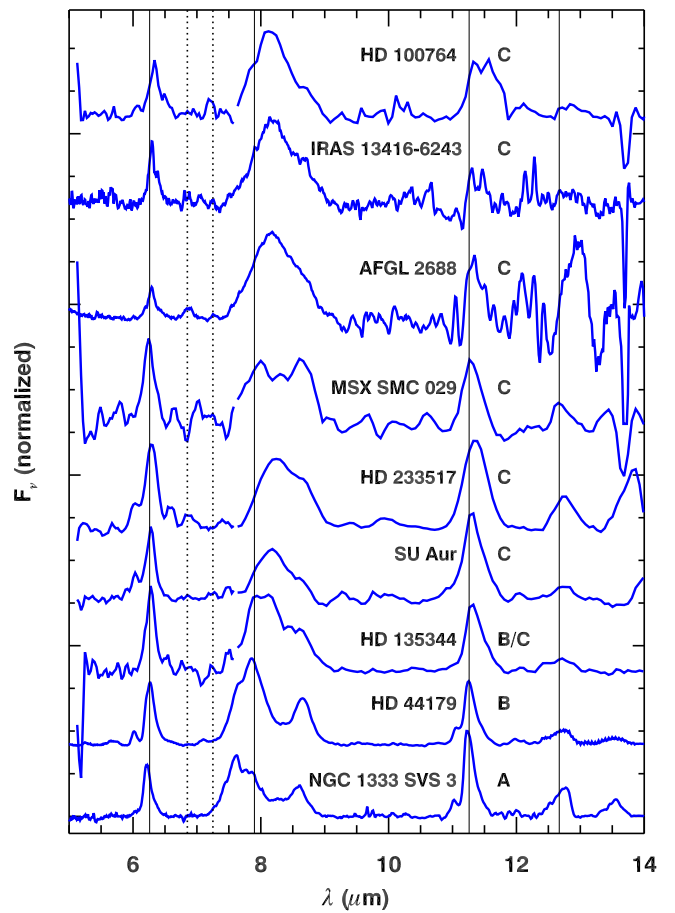}
\caption{The continuum-subtracted spectrum of HD 100764 
along with several comparison spectra (also
continuum-subtracted).  The name of each source is followed 
by the classification of its PAH spectrum.  The vertical 
lines are at the centers of the PAH features for a class B 
spectrum:  6.26, 7.90, 11.26, and 12.67~\mum.  The majority
of the class C spectra show features shifted to the red of
these lines.  The dashed lines at 6.85 and 7.25 mark the 
positions of aliphatic emission features.  The bottom two 
spectra are the prototypical class A and class B spectra from 
the \iso/SWS database.  The remainder are the sample of known 
class C PAH spectra (or in the case of HD 135344, nearly 
class C PAH spectrum).  The extractions of the PAH features 
at 12.0 and 12.7~\mum\ in AFGL 2688 are particularly 
untrustworthy due to their extremely low contrast and 
resultant sensitivity to the location of the spline points 
for the continuum.  In the spectrum of AFGL 2688, the bottom 
of the 13.7~\mum\ feature has been truncated to prevent it 
from overrunning the spectra below.}
\end{figure}

To place the class C PAH spectrum from HD 100764 in context,
we have also examined several other PAH spectra.  These 
comparison spectra include the two class C PAH sources from 
the SWS database, AFGL 2688 and IRAS 13416$-$6243, along with 
SWS spectra of NGC 1333 SVS 3 and HD 44179 (the Red 
Rectangle).  These sources are prototypical examples of class 
A and B PAH spectra, respectively.  All four spectra are from 
the atlas of SWS spectra produced by \cite{slo03}, and they 
have been regridded to a lower spectral resolution.  In the 
case of AFGL 2688, the spectrum is an average of two 
observations.  The comparison sample also includes the four 
sources observed by the IRS which are currently known to have 
PAH spectra either in or near class C, including the K giant 
HD 233517 \citep{jur06}, the post-AGB source MSX SMC 029 
\citep{kra06}, the HAeBe source HD 135344 \citep{slo05}, 
and the intermediate-mass T Tauri star SU Aur \citep{fur06}.  

Table 2 presents the comparison sample, including their 
spectral classes and effective temperatures, which are
described in \S 3.1.

% Table 2 goes here.

\subsection{Extracting the PAH features} % Sec. 2.3

\begin{deluxetable}{ccccc} % Table 3
\tablecolumns{5}
\tablewidth{0pt}
\tablenum{3}
\tablecaption{Wavelength intervals used to extract spectral features}
\tablehead{
  \colhead{} & \multicolumn{2}{c}{For class C sources} &
               \multicolumn{2}{c}{For class A and B sources} \\
  \colhead{Feature (\mum)} & 
  \colhead{$\lambda_{blue}$ (\mum)} & \colhead{$\lambda_{red}$ (\mum)} &
  \colhead{$\lambda_{blue}$ (\mum)} & \colhead{$\lambda_{red}$ (\mum)} }
\startdata
%6.0    &  5.89--5.95  &  6.07--6.13   & \\
 6.2    &  5.80--5.95  &  6.65--6.80   & as class C   & 6.54--6.69   \\
 7.9    &  7.30--7.60  &  9.00--9.30   & 7.10--7.40   & 8.92--9.17   \\
 8.6    &  8.32--8.44  &  8.86--8.98   & as class C   & as class C   \\
11.3    & 10.80--11.00 &  11.85--11.95 & 10.68--10.86 & 11.72--11.90 \\
12.0    & 11.85--11.95 &  12.35--12.60 & 11.72--11.90 & 12.13--12.25 \\
12.7    & 12.35--12.60 &  13.25--13.40 & 12.07--12.19 & 12.98--13.10 \\ 
13.7    & 13.25--13.55 &  13.83--14.13 & as class C   & as class C   \\
\enddata
\end{deluxetable}

\begin{center}
\begin{deluxetable*}{lccccccc} % Table 4
%\rotate
\tablecolumns{8}
\tablewidth{0pt}
\tablenum{4}
\tablecaption{Central wavelengths of the PAH and C$_2$H$_2$ features}
\tablehead{
  \colhead{ } & \multicolumn{6}{c}{$\lambda_{c}$ (\mum)} \\
  \colhead{Target} & \colhead{6.2~\mum\ PAH} & \colhead{7.7--8.2~\mum\ PAH} &
  \colhead{8.6~\mum\ PAH} & \colhead{11.3~\mum\ PAH} & 
  \colhead{12.0~\mum\ PAH} & \colhead{12.7~\mum\ PAH} & 
  \colhead{13.7~\mum\ C$_2$H$_2$} }
\startdata
HD 100764      &  6.33$\pm$0.01 &  8.15$\pm$0.06 & \nodata & 
                 11.47$\pm$0.07 & 12.14$\pm$0.15\tablenotemark{a} &
                 12.83$\pm$0.27\tablenotemark{a} & 13.70$\pm$0.02 \\
IRAS 13416     &  6.31$\pm$0.01 &  8.22$\pm$0.02 &  8.71$\pm$0.02 &
                 11.42$\pm$0.03 & 12.25$\pm$0.07 & 12.87$\pm$0.10 & 
                 13.70$\pm$0.00 \\
AFGL 2688      &  6.29$\pm$0.01 &  8.20$\pm$0.01 &  8.68$\pm$0.03 &
                 11.36$\pm$0.03 & 12.13$\pm$0.03 & 12.93$\pm$0.03 & 
                 13.71$\pm$0.01 \\
MSX SMC 029    &  6.25$\pm$0.04 &  8.12$\pm$0.10 &  8.68$\pm$0.05 &
                 11.32$\pm$0.07 & \nodata        & \nodata        & 
                 13.67$\pm$0.02 \\
HD 233517      &  6.28$\pm$0.03 &  8.26$\pm$0.03 &  8.72$\pm$0.07 &
                 11.35$\pm$0.02 & \nodata        & 12.75$\pm$0.09 & \nodata \\
SU Aur         &  6.26$\pm$0.01 &  8.19$\pm$0.03 &
                  8.67$\pm$0.18\tablenotemark{a} & 11.33$\pm$0.01 & 
                 12.02$\pm$0.03\tablenotemark{b} & 12.78$\pm$0.08 & \nodata \\
HD 135344      &  6.28$\pm$0.03 &  8.08$\pm$0.04 &  8.64$\pm$0.07 &
                 11.35$\pm$0.01 & \nodata        & 12.75$\pm$0.08 & \nodata \\
HD 44179       &  6.26$\pm$0.00 &  7.84$\pm$0.06 &  8.66$\pm$0.01 &
                 11.27$\pm$0.00 & 12.00$\pm$0.04 & 12.71$\pm$0.02 & \nodata\\
NGC 1333 SVS 3 &  6.23$\pm$0.00 &  7.68$\pm$0.01 &  8.61$\pm$0.01 &
                 11.26$\pm$0.01 & 12.00$\pm$0.02 & 12.70$\pm$0.01 & \nodata\\
\enddata
\tablenotetext{a}{S/N of extracted feature $<$ 2.}
\tablenotetext{b}{S/N of extracted feature $<$ 3.}
\end{deluxetable*}
\end{center}

\begin{center}
\begin{deluxetable*}{lccccccc} % Table 5
%\rotate
\tablecolumns{8}
\tablewidth{0pt}
\tablenum{5}
\tablecaption{Extracted strengths of the PAH and C$_2$H$_2$ features}
\tablehead{
  \colhead{ } & \multicolumn{6}{c}{$F$ (10$^{-15}$ W m$^2$)} &
  \colhead{$W_{\lambda}$ (10$^{-3}$ \mum) \tablenotemark{a}} \\
  \colhead{Target} & \colhead{6.2~\mum} & \colhead{7.7--8.2~\mum} &
  \colhead{8.6~~\mum} & \colhead{11.3~\mum} & \colhead{12.0~\mum} & 
  \colhead{12.7~\mum} & \colhead{13.7~\mum} } 
\startdata
HD 100764      &   4.61$\pm$0.24  &   9.15$\pm$0.82  &   \nodata       &
                   2.27$\pm$0.28  &   0.32$\pm$0.17  &   0.16$\pm$0.13 &
                   3.18$\pm$0.64 \\
IRAS 13416     &   57.3$\pm$2.4   &    225$\pm$5     &   1.73$\pm$0.51 &
                  22.9 $\pm$1.5   &   7.39$\pm$1.66  &   7.71$\pm$1.54 &
                   5.10$\pm$0.48 \\
AFGL 2688      &    111$\pm$4     &    860$\pm$10    &   18.6$\pm$1.6  &
                    147$\pm$15    &   76.5$\pm$7.4   &    204$\pm$14   &
                   6.23$\pm$0.57 \\
MSX SMC 029    &   0.16$\pm$0.03  &   0.29$\pm$0.03  & 0.051$\pm$0.007 &
                  0.068$\pm$0.012 &  0.005$\pm$0.021 & 0.001$\pm$0.015 &
                  14.84$\pm$2.96 \\
HD 233517      &   1.46$\pm$0.15  &   1.75$\pm$0.08  &  0.11$\pm$0.03  &
                   0.66$\pm$0.04  &   \nodata        & 0.091$\pm$0.037 &
                   1.22$\pm$3.09 \\
SU Aur         &   6.08$\pm$0.35  &  6.64$\pm$0.35   &   0.14$\pm$0.12 &
                   3.77$\pm$0.13  & 0.080$\pm$0.037  &   0.31$\pm$0.08 &
                   \nodata \\
HD 135344      &   5.44$\pm$0.54  &  9.42$\pm$0.46   &   0.43$\pm$0.12 &
                   2.34$\pm$0.06  &  \nodata         &   0.14$\pm$0.04 &
                   \nodata \\
HD 44179       &   2969$\pm$23    &   5731$\pm$677   &    960$\pm$19   &
                   1182$\pm$19    &   48.3$\pm$10.6  &    239$\pm$14   &
                   \nodata \\
NGC 1333 SVS 3 &    241$\pm$3     &    280$\pm$7     &   69.9$\pm$1.8  &
                    172$\pm$5     &   3.52$\pm$0.63  &   57.2$\pm$1.2  &  
                   \nodata \\
\enddata
\tablenotetext{a}{Equivalent width of the acetylene absorption band.}
\end{deluxetable*}
\end{center}

For each source, we follow the algorithm described for
HD 100764 by spline-fitting and removing a continuum,
as Figure 3 illustrates.  It has proven necessary to shift 
the spline points from one spectrum to the next to avoid 
some of the spectral structure which only appears in 
individual spectra.  Care must be taken in the case of 
IRAS 13416 and AFGL 2688 because the PAH features have such 
low contrast with respect to the continuum.  In the case of 
AFGL 2688, the features at 12.0 and 12.7~\mum\ must be 
treated with caution because their apparent strength and 
position shift substantially with the choice of spline 
points.  The spline fit to SU Aur removes the silicate 
emission feature; this method has been tested for several 
similar sources showing both PAH and silicate emission by 
\cite{kel07}.  Two of the spectra have been smoothed:
MSX SMC 029 with a 3-pixel boxcar, and HD 233517 with a 
5-pixel boxcar.  Figure 4 presents the continuum-subtracted 
PAH spectra for HD 100764 and the comparison sample.  

Fitting and removing a spline removes most, but not all, of 
the continuum spectrum.  Because we have not forced the fitted 
spline to pass precisely through the continuum on either side 
of each feature, small residual tilts remain.  While these
would have little impact on the extracted strengths of the
features, they would shift the measured center.  We therefore
continue to use the method applied to a sample of HAeBe
stars showing PAH emission by \cite{slo05}.  As they did, we 
fit line segments to the continuum on either side of each PAH 
feature and integrate the flux above this estimated continuum 
to measure the strength of the feature.  We determine the 
wavelength with half the emitted flux to either side, which
for the remainder of this paper, we call the central 
wavelength.

Other authors have analyzed the position of the peak 
wavelength, but this result differs from the central 
wavelength for two reasons.  First, PAH features are usually
asymmetric, so the peak will lie slightly to the blue of the
center.  Second, most of the features are blends of multiple
transitions.  The feature peaking at $\sim$7.7--7.9~\mum\ 
can be analyzed as a combination of features at 7.65 and 
7.85~\mum\ \citep[e.g.][]{pee02,bt05}.  As the 7.85~\mum\ 
component increases with respect to the 7.65~\mum\ component, 
the central wavelength will move steadily to the red, but the 
peak wavelength will not move significantly until the red 
component outshines the blue component.  Analyzing the 
central wavelength provides better insight about the relative 
strengths of the various components of each PAH feature.

% Table 3 goes here.

Table 3 gives the wavelength ranges used to fit the line 
segments to each feature; these have changed some from the 
ranges used by \cite{slo05}.  We use the class C 
wavelengths, for all of the sources except HD 44179 and 
NGC 1333 SVS 3, which require shifted wavelengths to extract 
the features.  All of the spectra except HD 100764 and SU Aur
have a feature at 8.6~\mum\ from the C--H in-plane bending
mode on the red shoulder of the C--C feature peaking between 
7.7 and 8.2~\mum.  To eliminate the influence of the 
8.6~\mum\ feature on the position and strength of the broader 
feature, we fit a line segment underneath it and include only
the flux below that line in our measurement of the 
7.7--8.2~\mum\ feature.

% Table 4 goes here.  
% Table 5 goes here.

Table 4 presents the extracted central wavelengths, while
Table 5 gives the equivalent fluxes (i.e., integrated flux 
densities), and for the acetylene band at 13.7~\mum, the 
equivalent width.  In cases where no feature is apparent 
or the strength of the extracted feature has a S/N ratio 
less than one, Table 4 does not quote a central wavelength.  
Table 4 includes notes indicating those wavelengths for 
features with S/N ratios less than 2 or 3.  Results for 
some of the comparison spectra have been published 
recently, but the results presented here represent newer 
S14 pipeline output and reflect modifications to the 
continuum fitting and extraction wavelengths.

\subsection{The PAH spectra} % Sec. 2.4

Figures 3 and 4 include the spectra of our class A and B
prototypes, NGC 1333 SVS 3 and HD 44179.  Using these and 
other SWS data, we find that the C--C modes produce features 
in class A spectra with typical central wavelengths of 6.23 
and 7.7~\mum, while class B spectra have features centered at 
6.26 and 7.9~\mum\ \citep[in essential agreement 
with][]{pee02}.  In the two class C spectra in the SWS 
database, these features are shifted to 6.30 and 8.2~\mum, 
respectively.  In HD 100764, they are centered at 6.33 and 
8.15~\mum.  In the other class C IRS sources, these features 
appear in the 6.25--6.28 and 8.09--8.29~\mum\ ranges.

The features beyond 10~\mum\ arise from out-of-plane C--H
bending modes.  The wavelength of these modes depends on the 
number of adjacent hydrogen atoms on a single aromatic ring
\citep[e.g.][]{atb89}.  Single H atoms produce the solo mode, 
which class A and B spectra show between 11.21 and 
11.30~\mum\ with a mean at 11.26~\mum.  Rings with two 
adjacent H atoms produce the duo mode, which normally appears 
at 12.00~\mum.  Three adjacent H atoms produce the trio mode 
which appears at 12.70~\mum\ in the class A and B spectra.  

In all of the class C sources, the solo out-of-plane bending
mode has shifted to longer wavelengths.  In the comparison
class C PAH spectra observed by the IRS, the features have
shifted from the nominal 11.26~\mum\ position to between 11.32 
and 11.36~\mum.  In the two class C spectra in the SWS database,
the shifts are difficult to measure due to possible effects from 
molecular absorption and artifacts.  Our best estimate places the 
band centers in the 11.36--11.42~\mum\ range.  In HD 100764 this 
feature is centered further to the red, at 11.47~\mum. 

The duo and trio modes also appear to shift in the class C 
spectra, but they are fainter and more difficult to measure.  
A shift is apparent in both features in the SWS spectra of 
AFGL 2688 and IRAS 13416, from 12.0 to $\sim$12.1--12.2~\mum\ 
for the duo mode and from 12.7 to $\sim$12.9~\mum\ for the 
trio mode, but these results should be treated with 
reservation.  The problem is that the contrast of these 
features is small, and their apparent strength and position 
depend on the details of how we fit a spline to the 
continuum.  Of the IRS comparison spectra, SU Aur shows the 
clearest duo-mode feature, but the S/N ratio is only $\sim$2 
and the shift is minimal.  The spectrum of HD 100764 shows a 
more significant shift, but again, the S/N ratio is only 
$\sim$2.  The situation with the trio mode is similar in the
IRS spectra.  A shift is apparent, but the uncertainties in 
central wavelength are large.

In summary, the positions of the PAH features in the spectrum
of HD 100764 are fully consistent with a class C PAH spectrum.
Two of the features, at 6.33 and 11.47~\mum, represent 
extreme positions for our sample.  The spectrum is definitely 
not class A or B.

\subsection{Other spectral features}  % Sec. 2.5

In Figures 2 and 4, the spectrum of HD 100764 shows structure
between the PAH features at 6.3 and 8.1~\mum\ reminiscent 
of the features observed in emission at 6.85 and 7.25~\mum\ 
in the spectra of the HAeBe stars HD 34282 and HD 169142 by 
\cite{slo05}.  These features are most obvious in the 
spectrum of AFGL 2688 (where they can be seen in both scan 
directions).  The carriers of the 6.85 and 7.25~\mum\ 
features have been identified as aliphatic hydrocarbons 
\citep{chi00} and/or hydrogenated amorphous carbon 
\citep[HAC;][]{fur99}.  In HD 100764, the features are faint 
and close to our uncertainty level.  The 7.25~\mum\ feature 
is stronger, which is a little unusual.  As a consequence, 
our identification of these features in HD 100764 is only 
tentative.  Other class C spectra also show possible spectral
structure in this area but with some variations.  For
example, IRAS 13416 shows an extra feature (all three appear 
in both scan directions).  The class A and B prototypes do 
not show the 6.85 and 7.25~\mum\ features.

Excluding the emission features from PAHs and related 
molecules, the remaining spectrum from HD 100764 consists of 
a cool dust continuum, absorption due to acetylene at 
13.7~\mum, and a broad emission feature centered at 
$\sim$10.6~\mum.  

The acetylene band is centered at 13.7~\mum, which is 
typical for a carbon star in the Galactic sample, and it has
an equivalent width of only 0.0032 $\pm$ 0.0006~\mum, about
a factor of 10--20 times weaker than seen in the SWS sample
of Galactic carbon stars \citep{slo06}.   The acetylene bands 
are apparent in sources with optically thick dust shells, 
indicating that the absorbing molecules are not in the 
stellar photosphere, but instead intermixed with or above the 
emitting dust \citep[][ and references within]{slo06}.  In the 
case of HD 100764, it is similarly likely that the acetylene 
absorption is not from the stellar photosphere.  Instead,
the acetylene is probably distributed in and/or above the 
disk.

The broad emission feature to the blue of the 11.47~\mum\
PAH feature can be fit reasonably well with a Gaussian
centered at 10.63~\mum\ with a full width at half maximum
of 1.37~\mum.  As \cite{spe05} have explained, typical SiC 
dust grains emit at $\sim$11.3~\mum, but larger SiC grains 
emit predominantly at 10.8~\mum\ because the contribution 
from the longitudinal optic at 10.8~\mum\ increases with 
respect to the transverse optic at 11.5~\mum\ as the grain 
size grows.  \cite{slo06} may have observed this effect in 
two carbon stars in the SMC, but they were trying to explain 
a shift from 11.3 to 11.0~\mum.  A shift past 10.8~\mum\ to 
10.6~\mum\ seems far less likely.  We conclude that this 
broad emission feature at 10.6~\mum\ is not likely to arise 
from SiC; it remains unidentified.

\section{Discussion} % Sec. 3.0

\subsection{Dependence of the PAH emission on excitation temperature} % Sec. 3.1

\begin{figure} % Fig. 5
\includegraphics[width=3.5in]{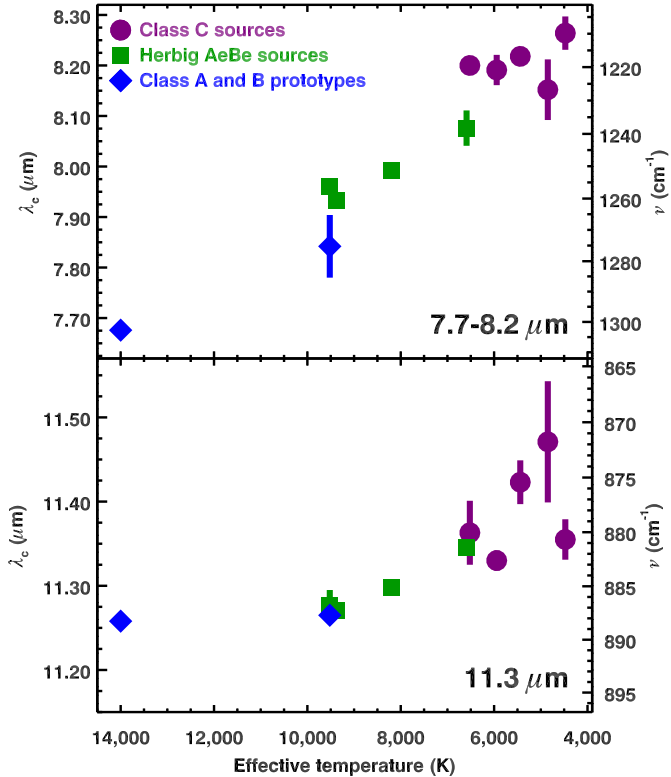}
\caption{The correlation between the central wavelength 
(defined in \S 2.3) of the two strongest PAH features and the 
effective temperature of the exciting star.  The top panel 
plots the position of the 7.7--8.2~\mum\ PAH complex, and the 
bottom panel plots the position of the 11.3~\mum\ feature.  
This figure contains three Herbig AeBe stars not included
in Tables 2, 4 or 5.  The IRS data for these sources, 
HD 34282, HD 141569, and HD 169142, appear elsewhere 
\citep{slo05,kel07}.  HD 135344 is the right-most square.  
The two prototypes are NGC 1333 SVS 3 (far left) and HD 44179 
(second to left).}
\end{figure}

Table 2 gives the spectral types and effective temperatures
of our PAH sample.  HD 100764 has been classified as R2 or 
C1,1 \citep{san44,yam72} and has an effective temperature of 
4850 K \citep{dom84}.  The two class C sources in the SWS 
database appear to be post-AGB objects evolving into 
planetary nebulae (PNe).  The central star in AFGL 2688 is 
classified as F5~Ia, and the surrounding nebulosity shows no 
strong emission lines \citep{cra75}.   IRAS 13416$-$6243 has 
a spectral type of G1~I \citep{hu93} and also shows no 
ionized emission \citep{kau93}\footnote{\cite{vds00} actually 
detected Br$\gamma$ {\it absorption} at 2.17~\mum.}.  
\cite{van89} describe it as in transition from the AGB to a 
PN.  We have determined temperatures from the spectral 
classes for these two objects using the relation calibrated 
by \cite{mcw91}.

The PAH spectrum of MSX SMC 029 is best described as class C.
While the 6.2~\mum\ feature has not shifted much from the 
position seen in class B, the 7--9~\mum\ PAH complex and 
11.3~\mum\ feature are shifted to the red.  Like the post-AGB
objects AFGL 2688 and IRAS 13416, MSX SMC 029 shows an 
absorption band from acetylene at 13.7~\mum.  \cite{kra06}
argue that it is one of the few known post-AGB objects in the 
SMC.  It is unfortunate that there are no optical data to help
classify the spectrum of the central source.

\cite{slo05} presented the spectra of four HAeBe stars, three 
classified as early to mid A dwarfs, and one, HD 135344, as 
F4 Ve \citep{dmr97}, which corresponds to an effective 
temperature of 6590 K \citep[using the calibration of][Table 
A5]{kh95}.  The PAHs around the A dwarfs resemble class B 
spectra, except that the 7.85~\mum\ feature is shifted 
to 7.9--8.0~\mum, and in the F dwarf HD 135344, this feature 
is shifted even further, to 8.08~\mum, making it intermediate
between class B and C.  \cite{kel07} have re-examined these 
four spectra as part of a larger sample of 18 HAeBe stars and 
four related stars, and they confirm these previous results.  
As a group, the HAeBe stars exhibit PAH spectra which differ 
from the typical class B PAH spectra, with more redshifted 
features.  The coolest stars in the sample have spectra which 
are even closer to class C.  One source examined by 
\cite{kel07} is the intermediate-mass T Tauri star SU Aur, 
which has a spectral class of G1 and an effective temperature
of 5945 K \citep{cal04}.  \cite{fur06} noted that while 
silicate emission dominates its infrared spectrum, emission 
from PAHs is apparent at 6.2~\mum\ and on the wings of the
10~\mum\ silicate feature at 8 and 11.3~\mum.  This source is 
the coolest star in the sample of \cite{kel07}, and it is 
their one source with an unambigious class C PAH spectrum.

HD 233517 has a spectral type of K2~III \citep{fek96}, an 
effective temperature of 4475 K \citep{bal00}, and in the
infrared, a class C PAH spectrum.  It is the coolest object 
in our comparison sample, and the center of its 7--9~\mum\ 
PAH complex is also at the longest wavelength observed.  

By comparison, the class A and B prototypes are warmer.
NGC 1333 SVS 3 has a spectral class of B6 \citep{hwj84},
which corresponds to an effective temperature of 14000 K
\citep{kh95}.  HD 44179 has a spectral type of A0 III
\citep{coh75}, corresponding to 9520 K \citep{kh95}.  While
a range of temperatures have been used to model the Red
Rectangle, we adopt 9520 K as most consistent with the
optical spectral class. %\cite{vih06} use a value of 8250 K.

Figure 5 plots the positions of the PAH features as a function
of the effective temperature of the exciting source.  In 
addition to the comparison sample in Tables 2, 4, and 5, 
the figure also includes the HAeBe stars HD 34282, HD 141569, 
and HD 169142.  \cite{slo05} first published the IRS spectra 
of these sources, and the results presented here are from the 
more recent analysis by \cite{kel07}.  The correlation of 
the position of the 7.7--8.2~\mum\ PAH complex with 
effective temperature is impressive.  While the 11.3~\mum\
PAH feature also tends to shift to longer wavelength as the
temperature decreases, the trend is not so clear, with little
difference between the class A and B sources and more
scatter in the class C sources.

\subsection{Spatial dependence of the PAHs in extended sources} % Sec. 3.2

In the original PAH model, ultraviolet radiation was required 
to excite the UIR emission features \citep[e.g.][]{atb89}.
Observations by \citet[][ 2000]{uch98} first observed PAH 
emission in cooler radiation fields, with effective 
temperatures as low as 6800 K in the case of the reflection
nebula vdB 133.  While these observations initially cast some 
doubt on the PAH model, \cite{ld02} showed that the absorption 
of visible photons by large or ionized PAHs can produce the 
observed UIR features.

Since PAH emission often arises in a photo-dissociation 
region (PDR), where a front is eating its way into a denser 
region and destroying the PAHs in the process, one might 
expect a change in the nature of the PAH spectrum as the 
temperature of the exciting radiation field varies.  
\cite{uch00} did not observe variation in the relative 
strengths of the features in their sample, but they did 
observe a change in the shape of the 7--9~\mum\ PAH complex.

\cite{bt05} analyzed the shape of the 7--9~\mum\ complex as 
a function of position within the reflection nebulae vdB 133, 
NGC 1333 SVS 3, and vdB 17 (also within NGC 1333).  They 
found that closer to the exciting star, the strength of the 
7.65~\mum\ component increases relative to the 7.85~\mum\ 
component, thus shifting the measured center of the 
7.7--7.9~\mum\ PAH feature to the blue.  They also found that 
this complex grows stronger relative to the C--H modes at 
longer wavelengths near the exciting sources, and suggested 
that the wavelength shift at 7.7--7.9~\mum\ might be related 
either to the ionization fraction of the PAHs or to PAH 
processing by the radiation field, as previously proposed 
by \cite{hon01}.  

%KK - raise distinction between UV hardness and UV intensity

\subsection{The possible role of aliphatics} % Sec. 3.3

Hydrocarbons can exist in both aromatic or aliphatic
forms.  Aromatic hydrocarbons are dominated by sp$^2$ bonds
which produce the polycyclic ring structure that define
PAHs.  Aliphatics include sp$^1$ bonds as seen in acetylene
and sp$^3$ bonds as seen in methane and longer alkane chains.
The sp$^2$ bonds efficiently spread the energy absorbed from
a photon over the entire molecule before it can disrupt a
single bond, making the aromatic carbon structure in PAHs 
more stable than aliphatic bonding structures.  Thus, when a 
mixture of aliphatic and amorphous hydrocarbons is exposed to 
a harsh radiation environment, we expect the aliphatic bonds 
to be broken first.  We propose that the class C spectra 
arise from mixtures of aromatic and aliphatic material that 
have not yet been subjected to intense ultraviolet radiation, 
allowing the more fragile aliphatic bonds to survive.

Consider the class C spectrum of MSX SMC 029.  This source
also shows a strong absorption band from acetylene 
(C$_2$H$_2$) at 13.7~\mum\ centered within a broader 
absorption band in the 12--16~\mum\ region \citep[see][Fig. 
4]{kra06}.  \cite{kra06} argue that this broader band 
is due to other aliphatic molecules seen in a similar 
spectrum at higher spectral resolution in SMP LMC 11, a 
source in Large Magellanic Cloud in transition to the early 
planetary nebula stage \citep[][Fig. 1]{jbs06}.  Acetylene 
appears in other class C PAH spectra, including AFGL 2688, 
IRAS 13146, and our spectrum of HD 100764\footnote{It is 
missing from HD 233517, which is an oxygen-rich source, and 
the disks around HD 135344 and SU Aur.}.  The association of
the acetylene absorption band at 13.7~\mum\ and the broader 
12--16~\mum\ absorption with class C PAH emission is 
consistent with our proposal that the class C spectra arise 
from mixtures of PAHs and aliphatics.

The ground-based mid-infrared spectrum of HD 56126 
(IRAS 07134+1005) shows red-shifted PAH features similar to 
our IRS spectrum of HD 100764, with emission both in a 
plateau at 8.0--8.2~\mum\ and from out-of-plane bending modes 
at 11.4 and 12.2~\mum\ \citep{jus96}.  Although not as 
pronounced, the infrared spectra of IRAS 04296+3429 and 
IRAS 05341+0852 display similar unusual characteristics.  
All three of these sources are post-AGB objects.  HD 56126 
is classified as F5 I \citep{nas65}, while IRAS 05341 and 
IRAS 04296 are F4 Iab: and G0 Ia, respectively.  These 
sources also show unusual spectra in the 3~\mum\ range, which 
is dominated by C--H stretching modes.  IRAS 05341 shows a 
feature at 3.40~\mum\ that dominates the normally stronger 
feature at 3.29~\mum\ \citep{gv90}, while IRAS 04296 also 
shows an unusually strong 3.40~\mum\ component 
\citep{geb92}.  The 3.29~\mum\ feature arises from an 
aromatic C--H stretching mode \citep[e.g.][]{dw81,atb89}, 
while the 3.4~\mum\ feature arises from a similar mode in 
aliphatic C--H bonds \citep{dw81,jm90,geb94}.  

In the Orion Bar, \cite{geb89} observed an increase in the 
ratio of the aliphatic 3.4~\mum\ feature to the aromatic 
3.29~\mum\ feature in the shielded molecular zone compared 
with the ionized zone.  \cite{slo97} verified these 
observations with long-slit 3~\mum\ spectroscopy and showed 
that the 3.4~\mum\ feature actually arises from multiple 
aliphatic components.  \cite{job96} obtained spatially 
resolved 3~\mum\ spectroscopy in the reflection nebulae 
NGC 1333 and NGC 2023 and found that where the ultraviolet 
radiation field was less intense, the 3.4~\mum\ feature is 
stronger compared to the 3.29~\mum\ feature.  All of these 
authors argued that the radiation field is altering an 
initial mix of aliphatic and aromatic bonds by destroying the 
more fragile aliphatic bonds.  

%(HD 56126 and IRAS 04296 also show a strong 21~\mum\ feature.)

\subsection{Hydrogenated amorphous carbon} % Sec. 3.4

\begin{figure} % Fig. 6
\includegraphics[width=3.5in]{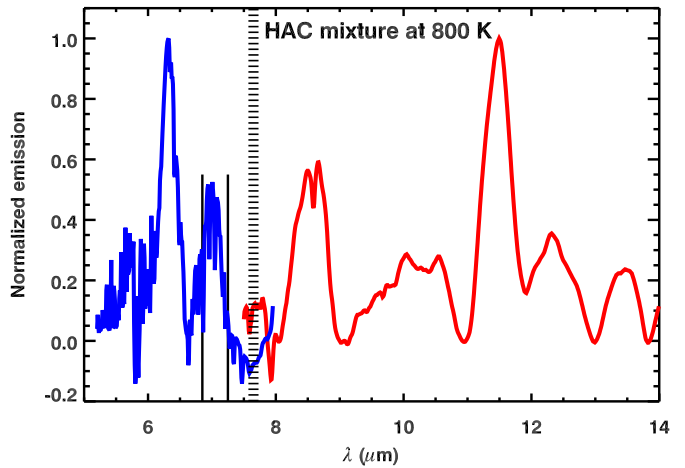}
\caption{A laboratory spectrum of hydrogenated amorphous 
carbon heated to 800 K\citep{sco97}.  The two portions of 
the spectrum separated by the vertical hatched line at 
7.5~\mum\ were obtained separately and have been arbitrarily 
normalized here to the same vertical range.  The apparent 
relative strengths of the two halves are not meaningful.  The
positions of the bands, however, are a good indication of
how the wavelengths of the various modes responsible
for the observed PAH features may shift as aliphatic bonds
are attached to the aromatic hydrocarbons that form the
basis of free PAH molecules.  Specifically, the two strongest
features at 6.32 and 11.45~\mum\ have shifted from 6.23 and
11.26~\mum\ in more normal PAH spectra.  The vertical lines
mark the positions of the 6.85 and 7.25~\mum\ features 
seen in spectra of cooler HAC samples.}
\end{figure}

Hydrogenated amorphous carbon (HAC) is a generic name for a 
mixture of aliphatic and aromatic carbon, consisting of 
PAH clusters embedded within a matrix of aliphatically bonded
material \citep[see ][ and references therein]{jdw90}.  This
description is similar to the material we are proposing to
explain the class C PAH spectra.

Compared to previously obtained class C PAH spectra, 
HD 100764 represents an extreme case because its features are 
more consistently red-shifted and its exciting radiation field 
is one of the coolest known.  The PAHs around this source must 
be relatively unprocessed by energetic photons, making this 
source an excellent candidate for the presence of HAC-like 
structures.

Figure 6 shows two representative laboratory spectra of a HAC 
sample at $\sim$ 800 K \citep{sco97}.  HAC films with a 
thickness of $\sim$1~\mum\ were prepared by ablating graphite 
in a hydrogen atmosphere with an excimer laser as reported 
previously \citep{sd96,sdj97}.  Films were deposited on a 
stainless steel substrate, placed in a vacuum chamber, and 
heated by conduction to $\sim$800 K.  Infrared emission 
spectra were obtained with a % Bomem MB-100 
Fourier-transform spectrometer at a resolution of 4 cm$^{-1}$ 
over a spectra range of 500--6000 cm$^{-1}$ (1.7--20~\mum).  
The spectra were calibrated by taking the ratio of the 
emission from HAC to that of an uncoated stainless steel 
substrate at the same temperatures and making a cubic 
polynomial baseline correction.  The two spectral segments 
from 5.5 to 8.0 and 7.5 to 14~\mum\ in Figure 6 were obtained 
separately and their relative intensities are unknown.  We 
have simply scaled both segments to a maximum of 1.0 
arbitrarily in the figure.  

While the relative band strengths in Figure 6 are not well
determined, the positions are more clear.  The emission 
features have shifted to the red, with the nominal 6.23 and 
11.26~\mum\ features appearing at 6.32 and 11.45~\mum, 
respectively.  If the features in the 12--13~\mum\ range can 
be attributed to the duo and trio C--H out-of-plane bending 
modes, then they have shifted from 11.99 and 12.67~\mum\ to 
12.39 and 13.43~\mum.  The feature centered at 8.53~\mum\ 
could be a red-shifted C--C mode, or it could simply be the 
C--H in-plane bending mode.  This laboratory HAC spectrum
is consistent with the wavelength shifts we have observed in
class C PAH sources.

\subsection{PAHs, HAC, and the class C spectrum} % Sec. 3.5.

\begin{figure} % Fig. 7
\includegraphics[width=3.5in]{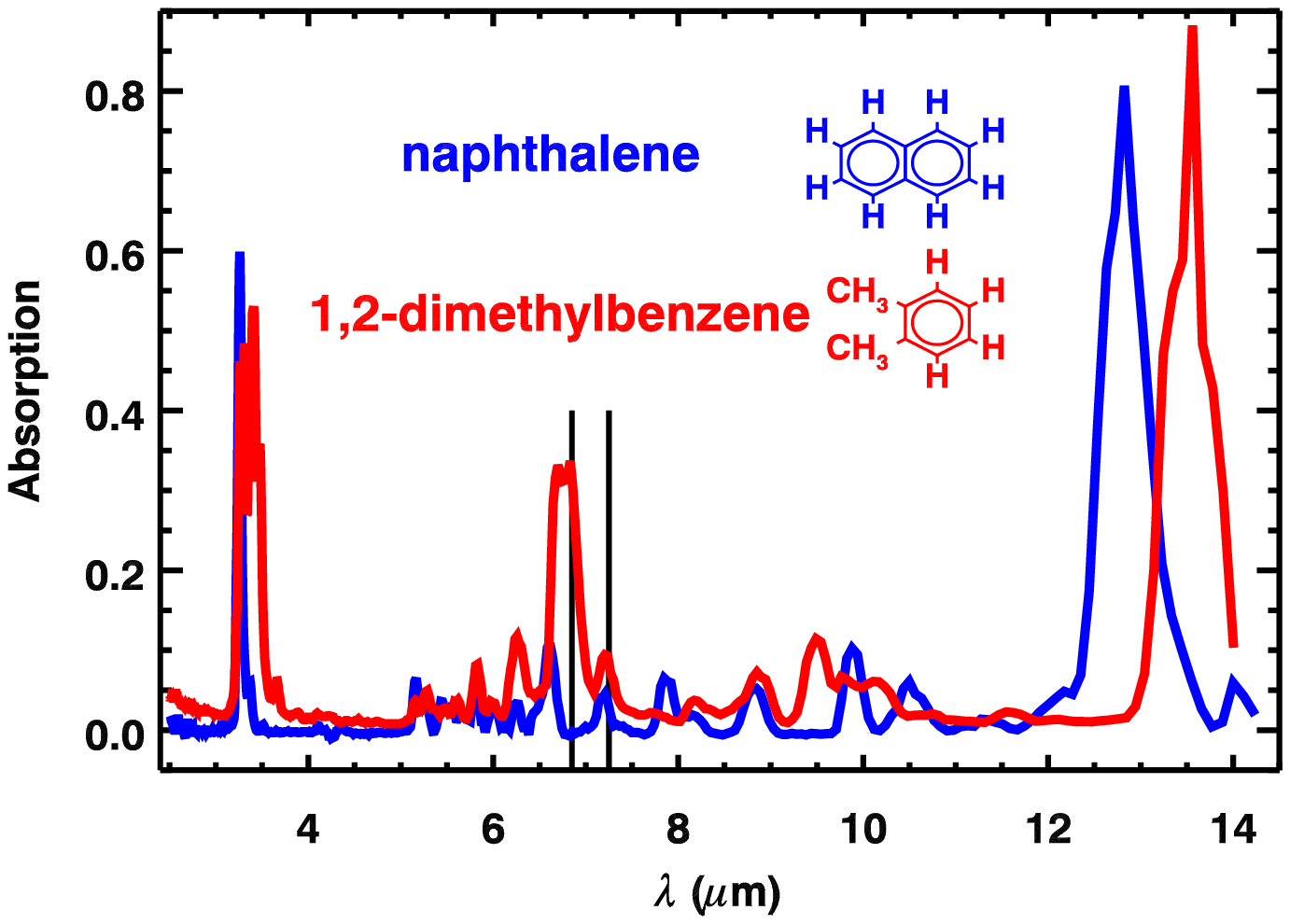}
\caption{Laboratory spectra of two simple PAHs:  naphthalene 
and benzene with two adjacent methyl side groups 
\citep[from][]{cob05}.  The subsitution of aliphatic 
sidegroups on benzene adds a second C--H stretching mode 
at 3.4~\mum, produces a strong absorption at 6.8~\mum, 
shifts the quartet out-of-plane C--H bending mode from 
12.8 to 13.5~\mum, and produces more spectral structure 
in the vicinity of the observed 6.85 and 7.25~\mum\
aliphatic features (indicated with solid vertical lines).}  
\end{figure}

Bonds in aromatic and aliphatic species have resonances at 
different wavelengths \citep[see][ and the many references 
therein]{bel78}.  In the case of the C--H stretching 
fundamental, the shift from aromatic to aliphatic is always 
to the red, from 3.29 to 3.40~\mum, but for other vibrational 
modes, the shifts can be in either direction, or both, 
depending on the molecules in question.  We hypothesize that 
at longer wavelengths, the substitution of aliphatic 
sidegroups on the edges of large PAH molecules shifts the 
out-of-plane aromatic bending modes to the red.  While 
laboratory data of large PAHs with aliphatic substitutions 
are lacking, data for some of the simplest PAHs are 
available, and as Figure 7 shows, are consistent with our 
hypothesis.

Using data from the NIST Chemistry WebBook, maintained by the
National Institute for Standards and Technology \footnote{At 
http://webbook.nist.gov/chemistry/}, we show in Figure 7 the 
spectra of naphthalene and 1,2-dimethylbenzene, which is 
benzene with two adjacent subsitutions of methyl sidegroups.  
The two spectra in Figure 7 were obtained from vapor samples 
using a Fourier-transform infrared spectrometer at a 
resolution of 4 cm$^{-1}$ \citep{cob05}.  The napthalene 
was at a temperature of 518 K, while the 1,2-dimethylbenzene 
was at 423 K.  Since both molecules have four adjacent H 
atoms on each ring, the only out-of-plane C--H bending mode
they show in the 11--14~\mum\ region is the quartet mode.
This mode has shifted from 12.8~\mum\ in naphthalene to 
13.5~\mum\ in the simpler benzene molecule with the methyl 
sidegroups, consistent with the anticipated effect of the  %%% REFERENCE?
substitution of a methyl sidegroup.  However, this spectral 
difference may also result from the absence of the second 
ring in naphthalene, making further analysis of larger 
molecules highly desirable.  

We also note that the presence of aliphatic C--H stretching 
modes in the methyl sidegroups on the benzene molecule has 
added emission at 3.4~\mum\ not present in the naphthalene.  
Furthermore, in the 6--8~\mum\ region, the strongest emisson 
feature in naphthalene is at 6.6~\mum, but a stronger feature 
at 6.8~\mum\ is apparent in the methyl-substituted benzene.  
As described by \cite{bel78} the carbon skeletal breathing 
modes in aromatics cluster in the 6.2 to 6.9~\mum\ range 
while similar modes from methyl (CH$_3$) and methylene 
(CH$_2$) groups cluster in the 6.8--7.5~\mum\ region.  As
noted in \S2.4, we may have detected these methyl and 
methylene features in emission in some of the class C 
PAH spectra, including HD 100764.

Laboratory HAC spectra by \cite{fur99} clearly show emission
features at 6.85 and 7.25~\mum.  The HAC spectrum in Figure 6
only shows a single emission feature at 7.0~\mum\ at about
half the strength of the 6.3~\mum\ feature.  The differences
in the spectra probably arise in how the samples were treated
and measured.  \cite{fur99} obtained absorption spectra from
samples at 300 K, while the HAC spectrum in Figure 6 was
observed in emission at $\sim$800 K.  This higher temperature 
has probably modified the simple groups responsible for the 
6.85 and 7.25~\mum\ features, and it has also enhanced the 
C$-$C modes at 6.2~\mum. 

% AL suggests we give the H/C ratio

HAC samples have a number of degrees of freedom, including 
the ratio of aliphatic and aromatic bonds and the size 
distribution and structure of the embedded PAHs among others, 
so it is not surprising that no HAC sample matches the 
observed astrophysical spectra in detail.  It is worth 
recalling that no laboratory PAH spectrum provides a precise 
match to astrophysical PAH spectra either.  Available data 
are consistent with the hypothesis that increasing the ratio 
of aliphatic to aromatic bonds in a hydrocarbon mixture 
explains the differences between class B and class C PAH 
spectra.  

We propose that complex astrophysical hydrocarbons are 
originally synthesized as some form of HAC.  When this 
material is exposed to harsh radiation fields, the aliphatic 
bonds are destroyed and only free PAHs remain.   \cite{sdp97}
provide evidence that this process could occur.  They exposed 
HAC mixtures to UV radiation and found that the dissolution 
of the HAC led to the production of large PAH molecules with 
attached aliphatic sidegroups.  \cite{hu06} have recently 
shown that these mixtures of aromatic and aliphatic bonds 
have the potential for reproducing the observed features 
in the 6--9~\mum\ range.  These modified PAHs have less than 
a few hundred atoms and still respond to the absorption of 
single photons like more typical free PAHs do.

Two separate processes are responsible for the PAH spectra
we observe.  First, the PAHs must be freed from the aliphatic
matrix in which they are initially embedded.  Second, the 
features must be excited.  This latter step requires less 
energy than that needed to completely strip the PAHs of all 
aliphatic side-chains.  The positions of the observed PAH 
features depend on how completely the PAHs have been freed 
from the underlying matrix.  In the class C PAH sources, the 
aliphatic side groups have only been partially destroyed.

The sample of isolated HAeBe stars represent a stage in the
evolution of PAHs between class C and class B.  The 
7.85~\mum\ feature is shifted measurably to 7.9--8.0~\mum\ or 
beyond \citep{slo05,kel07}.  Here again, the PAHs may not be 
entirely stripped from the underlying aliphatic matrix, since 
the A and F stars exciting the PAH spectrum may not produce 
enough photons with sufficient energy to break all of the 
aliphatic bonds.  

\subsection{The problem of the reflection nebulae} % Sec. 3.6

RNe typically do not show class C PAH spectra, even though
the exciting source can be as cool as 6800 K \citep{uch98}.
The two RNe in the sample examined by \cite{pee02} are both 
class A, as are the three spectra presented by \cite{uch00}.  
The objects studied by \cite{bt05} include one from 
Uchida et al., and all three showed variations in the 
position of the 7.7--7.9~\mum\ feature with radiation 
field, but not enough to move them even as far as a true 
class B spectrum.

We suggest that while the PAHs in these complex, evolving 
environments are now subject only to cool radiation fields, 
they may have been exposed in the recent past to a large 
dose of more intense ultraviolet radiation.  RNe are 
typically diffuse and are associated with complicated
star-forming environments.  This diffuse nature means that
the PAHs are poorly shielded from both current and past 
radiation fields.  Furthermore, PAHs in a diffuse environment 
are not able to regenerate any broken aliphatic bands. 

%While reflection nebulae may be the strongest example of
%the interplay of the processing and excitation mechanisms 
%necessary for PAH emission, it is unlikely that they are the
%only example.  
The typical PAH emission source is a PDR in a star formation 
region, and the emitting materials in these environments must 
have a complicated history of UV irradiation.  This history 
blurs the link between the current PAH composition and the 
current radiation field, which is probably why the 
correlation illustrated in Fig. 5 has not been noticed 
before.  Only in more isolated sources such as those examined 
here is the link clear.

\section{Conclusions} % Sec. 4.0

The PAH spectrum of HD 100764 adds to the small, but growing
sample of class C PAH spectra, as defined by \cite{pee02}.  
It has one of the coolest exciting stars and some of the
most redshifted features in the class.  Since all 
seven of the known class C spectra are excited by relatively 
cool stars of spectral class F or later, we argue that the
hydrocarbons in these systems are unusual because they have 
not been exposed to much ultraviolet radiation.  

We hypothesize that carbonaceous materials are synthesized as 
large HAC conglomerates, and that aliphatic bonds are 
subsequently broken in harsh radiation fields.  We suggest 
that PAH spectra evolve over time; the hydrocarbons in 
class C objects are relatively protected and unprocessed 
while class B and A PAHs have been exposed to more energetic 
photons and are more processed.

\acknowledgments

We are grateful to the referee, whose careful reading and
thoughtful comments led to considerable improvements in the
manuscript.  We would also like to thank T. Geballe and 
P. Roche for useful comments and suggestions and M. Werner 
for generously providing us with his high-resolution spectra 
of HD 100764.  These observations were made with the {\it 
Spitzer Space Telescope}, which is operated by JPL, California 
Institute of Technology, under NASA contract 1407 and 
supported by NASA through JPL (contract 1257184).  Support for 
this work was provided by NASA through contract 960803 issued 
by JPL/Caltech.  WMD is supported by a grant from the NSERC 
of Canada.  AL is supported in part by the NASA/{\it Spitzer} 
theory program.  This research has made use of the SIMBAD 
database and VizieR tool for catalog access, operated at the 
Centre de Données astronomiques de Strasbourg.

\end{document}